\documentclass[aps,prb,reprint,twocolumn]{revtex4-2}
\usepackage{inputenc,amsfonts,amsmath,amsthm,amssymb,amsmath,amscd,graphicx,mathrsfs,braket,hyperref,xcolor,dsfont,physics,comment}

\renewcommand\vec{\boldsymbol}
\def\r{{\vec{r}}}
\def\x{{\vec{x}}}
\def\y{{\vec{y}}}
\def\k{{\vec{k}}}
\def\p{{\vec{p}}}

\def\q{{\vec{q}}}
\def\A{{\vec{A}}}
\def\T{{\vec{T}}}

\def\th{{\vec\theta}}

\def\C{\mathbb{C}}
\def\al{\alpha}
\def\A{\mathbb{A}}

\begin{document}
\title{Area-law entanglement from quantum geometry}
\author{Nisarga Paul}
\email{npaul@mit.edu}
\affiliation{Department of Physics, Massachusetts Institute of Technology, Cambridge, MA,
USA}

\begin{abstract}
Quantum geometry, which encompasses both Berry curvature and the quantum metric, plays a key role in multi-band interacting electron systems. We study the entanglement entropy of a region of linear size $\ell$ in fermion systems with nontrivial quantum geometry, i.e. whose Bloch states have nontrivial $k$ dependence. We show that the entanglement entropy scales as $S = \alpha \ell^{d-1} \ln\ell + \beta  \ell^{d-1} + \cdots$ where the first term is the well-known area-law violating term for fermions and $\beta$ contains the leading contribution from quantum geometry. We compute this for the case of uniform quantum geometry and cubic domains and provide numerical results for the Su-Schrieffer-Heeger model, 2D massive Dirac cone, and 2D Chern bands. An experimental probe of the quantum geometric entanglement entropy is proposed using particle number fluctuations. We offer an intuitive account of the area-law entanglement related to the spread of maximally localized Wannier functions. \end{abstract}

\maketitle

\section{Introduction}
A central topic in current condensed matter physics is the study of strongly correlated phases in partially filled Chern bands, such as unconventional superconductivity, Mott insulators, and fractional Chern insulators \cite{Cao2018Apr,Cao2018Apr2,Li2021Dec,Spanton2018Apr}. For non-topological bands, the study of strongly correlated phases often begins with the Hubbard model, since one can always find a basis of exponentially localized Wannier orbitals. In contrast, one cannot find such a basis for topological bands in greater than one dimension, a fact known as the Wannier obstruction \cite{Brouder2007Jan}. For example, in a Landau level in a disordered potential, most states are localized while a narrow band of states must be delocalized-- which is crucial for the integer quantum Hall effect. \par 

The inability to find exponentially localized Wannier orbitals in topological bands should imply longer-range entanglement. Entanglement has proven a valuable conceptual tool and diagnostic in topological condensed matter physics \cite{Kitaev2006Mar,Zeng2015Aug,Li2008Jul}, but there has not been a detailed study of real-space entanglement in partially filled Chern bands. 
\par
We find that an essential quantity is the quantum geometry of the partially filled band. Quantum geometry is simply the structure of Bloch state overlaps, and it encompasses both Berry curvature and the quantum metric \cite{Provost1980Sep}. Recent works have highlighted the importance of the full quantum geometry, and not simply Berry curvature, in the study of interactions in Chern bands, for instance in twisted bilayer graphene \cite{Parameswaran2013Feb,Ledwith2020May,Parker2021Dec}. 
\par

In particular we show that nontrivial quantum geometry contributes additional area-law entanglement; that is, the von Neumann entropy of a region of linear size $\ell$ in $d$ dimensional systems goes as 
\begin{equation}\label{eq:S}
    S = \alpha \ell^{d-1} \ln \ell + \beta \ell^{d-1} + \cdots
\end{equation}
where the quantum geometry contributes at order $\ell^{d-1}$. The leading term is a log-enhanced area law originating from the $O(\ell^{d-1})$ number of independent 1d chiral modes on the Fermi surface, each of which contributes $\ln \ell$ \cite{SwingleJul2010}. $\alpha$ can be computed using Widom's conjecture and depends only on the entangling region and Fermi surface, not on the quantum geometry \cite{Gioev2006Mar}.
\par 

We find it useful to define the \textit{quantum geometric entanglement entropy} $S_{QG}$ by subtracting from $S$ the entropy obtained by setting all Bloch overlaps $\left \langle u_\k | u_\q\right \rangle$ to unity. In other words, 
\begin{equation}
    S_{QG} = S- S|_{\left\langle u_\k|u_\q\right\rangle \to 1}.
\end{equation}
$S_{QG}$ measures the additional entropy coming from intrinsically multi-band effects, omitting contributions shared with continuum fermions with an identical Fermi surface. \par

Our main results can then be expressed as follows: 

\begin{enumerate}
    \item We show $S_{QG} = \beta \ell^{d-1}+\cdots$ for the case of uniform quantum geometry (i.e. $\left\langle u_\k|u_{\k+\q}\right\rangle = g(\q)$) and provide a closed form for $\beta$
    \item We show numerically that $S_{QG}$ is area law for the 1D Su-Schrieffer-Heeger model, the 2D massive Dirac fermion, and 2D Chern bands
    \item We establish that particle number variance receives a similar area-law contribution from quantum geometry and propose an experimental measure of $S_{QG}$
\end{enumerate}
As a corollary to 3, we find a protocol to measure the quantum metric at the band bottom using particle number fluctuations. \par 
This paper is organized as follows. In Sec. \ref{sec:prelim} we define fermion entanglement entropy and other preliminaries. In Sec. \ref{sec:simp} we provide a simple two-particle example which captures general features of quantum geometric entanglement entropy. In Sections \ref{sec:simp}, \ref{sec:ssh}, and \ref{sec:2d} we calculate $S_{QG}$ for the cases of uniform quantum geometry in $d$ dimensions, the 1d Su-Schrieffer-Heeger model, and 2d models, respectively. In Sec. \ref{sec:exp} we establish similar results for particle number fluctuations and suggest an experimental protocol for measuring $S_{QG}$. In Sec. \ref{sec:wannier} we provide an intuitive explanation for the area-law behavior of $S_{QG}$ and its relation to quantum geometry using Wannier orbital spread, and in Sec. \ref{sec:discussion} we conclude.

\section{Preliminaries}
\label{sec:prelim}

In this section we review how entanglement entropy can be computed for fermions, although we note for the reader that only Eq. \eqref{eq:Smain} is needed in the sections that follow. We start by considering a system of $N$ noninteracting fermions with a discrete one-particle spectrum. The many-body wavefunction takes the form $\Psi(\x_1,\ldots, \x_N) = \det[\phi_n(\x_i)]/\sqrt{N!}$ where $\x_i$ includes both spatial and internal coordinates (e.g. spin / sublattice) and the $\phi_n$ are the normalized wavefunctions of the occupied states. The partial density matrix of a spatial region $A$ is 
\begin{equation}
    \rho_A = \Tr_{\bar{A}} \ket{\Psi}\bra{\Psi}
\end{equation}
and the $\alpha$'th R\'enyi entropy is
\begin{equation}
    S^{(\alpha)}(A) = \frac{1}{1-\alpha} \ln \Tr \rho_A^{\alpha},
\end{equation}
with $\alpha \to 1$ the von Neumann entropy. We have the useful relation \cite{Peschel2003Mar} 
\begin{equation}
    \rho_A = (\det \C )\exp\left[\sum_{\x,\y \in A} \ln\left(\C^{-1}-\mathbb{I}\right)_{\x,\y} c_\x^\dagger c_\y\right]
\end{equation}
where
\begin{equation}
    \C(\x,\y) = \left\langle c^\dagger(\x) c(\y)\right\rangle = \sum_{n=1}^N\phi_n^*(\x)\phi_n(\y)
\end{equation}
is the two-point correlator in $\Psi$ restricted to $A$. This allows us to express the R\'enyi entropies as
\begin{equation}
    S^{(\alpha)}(A) = \sum_\ell e_\alpha (\lambda_\ell) 
\end{equation}
where
\begin{equation}
    e_\alpha(\lambda) = \frac{1}{1-\alpha} \ln(\lambda^\alpha +(1-\lambda)^\alpha)
\end{equation}
and $\lambda_\ell$ are the eigenvalues of $\C$. Note that the limit $\alpha\to 1$ gives
\begin{equation}
    e_1(\lambda) = -\lambda \ln \lambda - (1-\lambda)\ln (1-\lambda).
\end{equation}
Since $\C$ grows in size with $A$, computing its spectrum can be a nontrivial task even on a lattice. To simplify things, we use the Fredholm determinant 
\begin{equation}
    D_A(\lambda) = \det[\lambda \mathbb{I} - \mathbb{C}]
\end{equation}
to rewrite $S^{(\alpha)}(A)$ as  \cite{Jin2004Aug,Calabrese2011Jul}
\begin{equation}\label{eq:contour}
    S^{(\alpha)}(A) = \oint \frac{d\lambda}{2\pi i} e_\alpha(\lambda) \dv{\ln D_A(\lambda)}{\lambda}
\end{equation}
where the integration contour encircles the segment $[0,1]$. Next we introduce the overlap matrix \cite{Klich2006Jan,Calabrese2011Jul}
\begin{equation}\label{eq:overlap}
    \A_{nm} = \int_A d\x \,\, \phi_n^*(\x) \phi_m(\x).
\end{equation}
Importantly, by $\int_A d\x$ we mean an integral over the spatial region $A$ \textit{and} a sum over internal indices. This matrix enjoys the property 
\begin{align}
    &\Tr \A^k = \nonumber\\
    &\sum_{n_1,\ldots, n_k=1}^N\int_A d\x_1  \phi_{n_1}^*(\x_1) \phi_{n_2}(\x_1) \cdots \int_A d\x_k\phi_{n_k}^*(\x_k)\phi_{n_1}(\x_k)\nonumber\\
    &= \int_A d\x_1 \cdots \int_A d\x_k \C(\x_1,\x_k) \C(\x_k,\x_{k-1})\cdots \C(\x_2,\x_1)\nonumber\\
    &\quad\qquad= \Tr \C^k
\end{align}
and therefore \cite{Calabrese2011Jul}
\begin{align}
    \ln D_A(\lambda) &= \ln\lambda-\sum_{k=1}^{\infty}\frac{\Tr \C^k}{k\lambda^k} = \ln\lambda-\sum_{k=1}^{\infty} \frac{\Tr \A^k}{k\lambda^k}\nonumber\\
    &= \sum_{m=1}^N\ln(\lambda-a_m)
\end{align}
where the $a_m$ are the eigenvalues of $\A$. We apply the residue theorem to Eq. \eqref{eq:contour} to obtain
\begin{equation}\label{eq:Smain}
    S^{(\alpha)}(A) = \sum_m e_\alpha(a_m),
\end{equation}
which lends itself well to numerical calculations since $\A$ is an $N\times N$ matrix, independent of the entangling region.

\section{Simple two-particle example}

\label{sec:simp}

Some implications of quantum geometry for entanglement entropy can be seen in a simple two-particle example. Consider a Slater determinant of two occupied single-particle states in a translationally invariant $m$-band system,
\begin{equation}
    \phi_i(\r) =L^{-d/2} e^{i \k_i \cdot \r} \ket{ u_{\k_i}},\quad i=1,2,
\end{equation}
where $L^d$ is the system size and $\ket{ u_{\k_i}}$ is an $m$-component vector. We will parameterize the Bloch overlap as
\begin{equation}
   \left\langle u_{\k_1}|u_{\k_2}\right\rangle = u e^{i\phi}. 
\end{equation}
When $u=1$, the quantum geometry is trivial ($\phi$ can be gauged away). Therefore the deviation of $u$ from $1$ is a measure of quantum geometry, in a way we make precise. Let the entangling region be $A$. The overlap matrix is
\begin{equation}
    \mathbb A = \begin{pmatrix} I_0 & I_{d\k} ue^{i\phi}\\
    I_{-d\k} ue^{-i\phi} & I_0
    \end{pmatrix}
\end{equation}
where $d\k \equiv \k_2-\k_1$ and
\begin{equation}
    I_\k = L^{-d}\int_A d^d\r \, e^{-i\k \cdot \r}.
\end{equation}
The eigenvalues of the overlap matrix are 
\begin{equation}
    a_{1,2} = I_0 \pm |I_{d\k}| u
\end{equation}
and the entropy is $S^{(\al)} = \sum_i e_\al (a_i)$. It is easy to check that $\left.\partial S^{(\al)}/\partial u\right|_{u=1} <0$, so that $S^{(\al)}$ increases as $u$ decreases from $1$. In fact, since $e_\al$ is Schur-concave, $S^{(\al)}$ increases monotonically with decreasing $u$. The entropy is maximized when $u=0$, i.e. when the Bloch states are orthogonal. \par 
In this example, the entanglement entropy ``increases with quantum geometry." We can make this more precise if $d\k \equiv \k_2-\k_1$ is small, in which case
\begin{equation}
    1- u^2 \approx g_{\mu\nu}(\k_1) dk_\mu dk_\nu
\end{equation}
where $g_{\mu\nu}(\k)$ is the quantum metric. Then the entanglement entropy increases with the quantum metric.\par 

In general, nontrivial quantum geometry has the effect of decreasing the magnitudes of  off-diagonals of $\mathbb A$. Since the trace is unaffected, this drives the eigenvalues towards their average value, increasing the entropy. While the phase $\phi$ of the Bloch overlap played no role in this simple two-state example, phases play a role as soon as three or more states are involved, and in general the phase structure is captured by Berry curvature.

\section{Uniform quantum geometry}

\label{sec:unif}

\begin{figure}
    \centering
\includegraphics[width=0.9\linewidth]{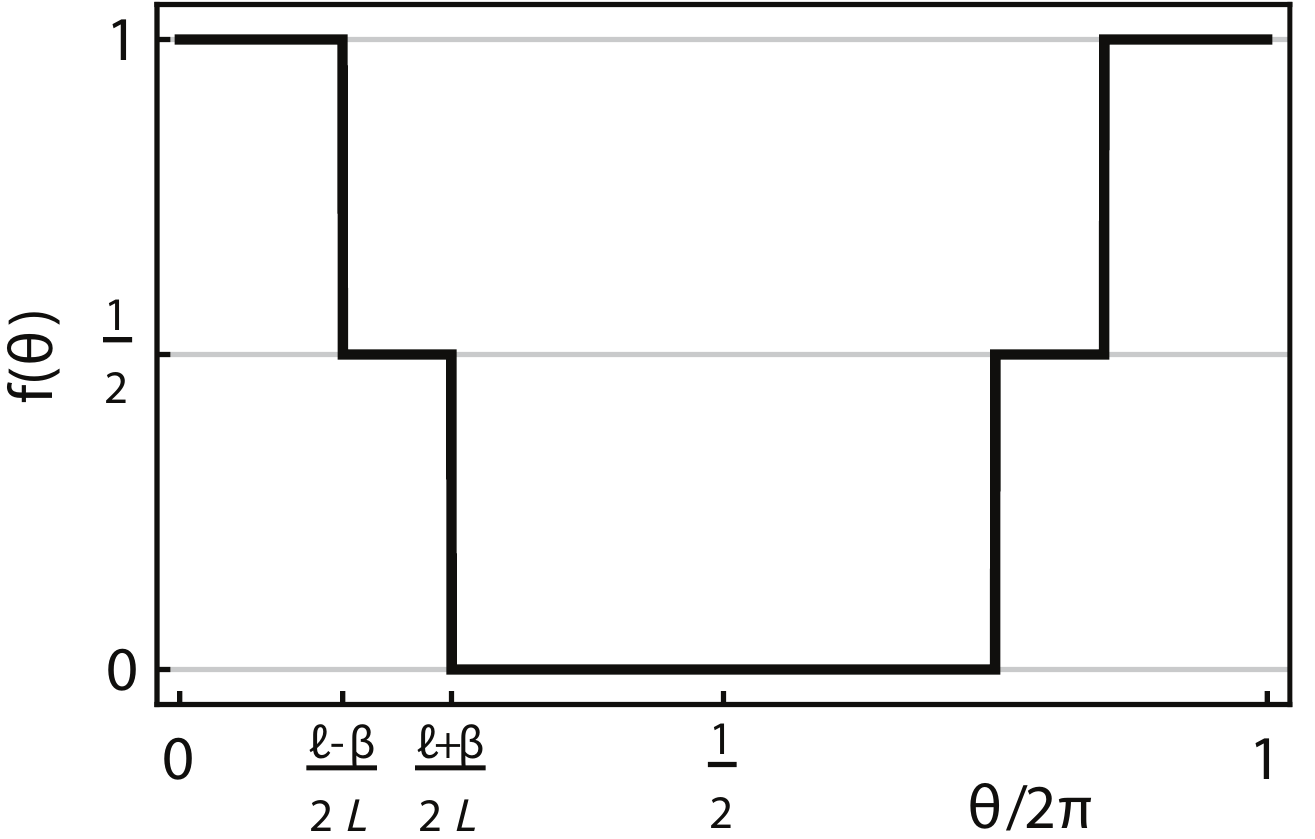}
    \caption{Plot of $f(\theta)$ (from Eq. \eqref{eq:sym}). Note that $e_1(0)=e_1(1)=0$ so only the portion where $f(\theta)=1/2$ contributes, and hence $S_{QG} \sim \beta$.}
    \label{fig:symbol}
\end{figure}

Before generalizing to $d$ dimensions, let us first consider a two-band model in a 1d system of size $L$ with periodic boundary conditions. As a simple model with quantum geometry, we take the wavefunctions
\begin{align}\label{eq:wvfns}
    \phi_{1,k}(x) &= L^{-1/2} e^{ikx} \begin{pmatrix} \cos(\beta k)\\\sin(\beta k)\end{pmatrix}\\
    \phi_{2,k}(x) &= L^{-1/2} e^{ikx}\begin{pmatrix} -\sin(\beta k)\\\cos(\beta k)\end{pmatrix} 
\end{align}
for the 1st and 2nd band, respectively, and $k \in 2\pi \mathbb{Z}/L$. Let $A = [-\ell/2,\ell/2]$ be the entangling region and assume the 1st band is filled in the region $\Gamma = [-k_F,k_F]$. The overlap matrix is
\begin{align}
    \mathbb{A}_{k,k'} &= \int_A dx\,  \phi^*_{1,k}(x)\cdot \phi_{1,k'}(x)\nonumber\\
    &= L^{-1}\int_A dx\, e^{- i(k-k')x}  \cos(\beta (k-k'))\nonumber\\
    &= \frac{2 \sin ((k-k')\ell/2)}{L(k-k')}\cos(\beta(k-k'))\\
    &\equiv f(k-k')\nonumber
\end{align}
Since the matrix elements depend only on $k-k'$, $\mathbb{A}$ is a Toeplitz matrix. Its spectrum converges at large $N$ to the image of its ``symbol"
\begin{align}
    f(\theta) &= \sum_{k} f(k) e^{ikL \theta/2\pi} \nonumber\\
    &= \frac12 \sum_{s=\pm} \sum_k \frac{2 \sin k \ell / 2}{kL} e^{ik(\frac{L}{2\pi} \theta + s\beta)}\nonumber\\
    &= \frac12 \sum_{s=\pm} (1-\chi_{[\frac{\ell}{2L}, 1-\frac{\ell}{2L}]}( \theta/2\pi + s \beta/L))\label{eq:sym}
\end{align}
where $\chi_{[a,b]}(x) = 1$ if $ x\in [a,b]$ and $0$ otherwise. This is plotted in Fig. \ref{fig:symbol}, restricted to $0 \leq \theta<2\pi $. Next, we use the classical Szeg{\H o} theorem to compute \cite{grenander_szego} 
\begin{align}
    S_{QG} &= \frac{N}{2\pi} \int_0^{2\pi} e_1(f(\theta)) \,d\theta + \cdots \nonumber \\
    &= 2(k_F/\pi) |\beta|  \ln 2 + \cdots
\end{align}
where we've used $2k_F = 2\pi N / L$ and the $\cdots$ are subleading terms going as $\ell^{-1}\ln \ell$ which can be calculated analytically using the Fisher-Hartwig theorem. Each boundary point contributes $\sim\ln 2$, resulting in an area-law contribution to entanglement entropy. \par  

To be concrete, we could provide a lattice model with the eigenfunctions in Eq. \ref{eq:wvfns} even though the particulars of the model aren't needed to compute the entanglement entropy. Let's choose band energies $\epsilon_{1,2}(k) = \mp (t_0 + t_1 \cos ka)$ for spinless fermions on a linear chain with two sublattices $A$ and $B$. The model is

\begin{multline}\label{eq:lattice}
    H = \sum_j \left[\sum_{\delta} \frac{t_\delta}{2} (c_{Aj}^\dagger c_{A,j+\delta} - c_{Bj}^\dagger c_{B,j+\delta}) \right.\\\left.+i\sum_{\delta'} t_{\delta'} c_{Aj}^\dagger c_{B,j+\delta'} -i\sum_{\delta''}t_{\delta''} c_{Aj}^\dagger c_{B,j+\delta''} + \text{h.c.}\right]
\end{multline}
where  $\delta = \pm 2\beta/a, \pm 1 \pm 2\beta/a$, $\delta'=+2\beta/a, \pm 1+ 2\beta/a$, $\delta''=-2\beta/a, \pm 1 -2\beta/a$, $t_{\pm 2\beta/a} = -t_0/2$ and $t_{\pm 1 \pm 2\beta/a} = -t_1/4$, and $2\beta /a \in \mathbb{Z}$.
\par

This explicit lattice model suggests a natural interpretation for $S_{QG}$ in this 1D example. The lattice model involves hopping by a range proportional to $\beta$, which results in an area-law increase in entanglement entropy proportional to $\beta$. We return to this point in Sec. \ref{sec:wannier}. \par 

Next we show that the entanglement entropy is area-law in $d$ dimensions. In this case we work with a multilevel Toeplitz matrix, and the classical Szeg{\H o} theorem generalizes straightforwardly \cite{Tyrtyshnikov1998Feb}. We will work with cubic regions $A = [-\ell/2, \ell/2]^d$ for convenience, though the leading behavior at large $\ell$ should depend only weakly on this choice. \par

The symbol is now a function of a vector $\th$ of $d$ angles. We can write
\begin{equation}
    f(\th) = \sum_\k \left(\prod_{i=1}^d \frac{2\sin k_{i} \ell/2}{k_{i}L} \right)
  \cos(\vec\beta\cdot \k) e^{i(L/2\pi)\k \cdot \th}
\end{equation}
where we've generalized to a vector $\vec \beta$ and the sum is over $(\frac{2\pi}{L}\mathbb{Z})^d$. This straightforwardly gives
\begin{equation}
    f(\th) = \frac12 \sum_{s=\pm } \prod_{i=1}^d (1-\chi_{[\frac{\ell}{2L}, 1-\frac{\ell}{2L}]}( \theta_i/2\pi + s \beta_i/L)).
\end{equation}
The entropy is given by 
\begin{align}
   S_{QG} &= \left(\frac{N}{2\pi}\right)^d \int e_1(f(\th)) \, d^d\th+ \cdots \nonumber\\
    &= 2 (k_F/\pi)^d\ell^{d-1}\sum_{i=1}^d |\beta_i| \ln 2 + \cdots\label{eq:ld-1}
\end{align}
where the $\cdots$ are subleading in $\ell$.

Next we treat the general case with uniform quantum geometry. By this we mean 
\begin{equation}
\left\langle u_\q | u_{\q + \k}\right\rangle 
 = g(\k),
 \end{equation}
 i.e. Bloch overlaps depend only on the momentum difference $\k$. Since $g(- \k) = g( \k)^*$, we may write this in the continuum setting as
 \begin{equation}
     g(\k) = \int d^d \vec \beta \, \tilde g(\vec \beta) e^{i\vec \beta \cdot \k}
 \end{equation}
where $\tilde g(\vec \beta)$ is a real-valued density integrating to 1. It is useful to define the marginal density of $\beta_i$ as 

\begin{equation}
    \tilde g_i(\gamma) = \int (\prod_{j\neq i} d\beta_j )\, \tilde g(\vec \beta)|_{\beta_i = \gamma}
\end{equation}
and the cumulative marginal density as 
\begin{equation}
    \tilde G_i(\gamma) = \int_{-\infty}^\gamma d\gamma'\, \tilde g_i(\gamma').
\end{equation}
Applying the classical Szeg{\H o} theorem then gives
\begin{equation}
    S_{QG} = \left(\frac{k_F}{\pi}\right)^d \ell^{d-1}\sum_{i=1}^d \int d\beta  \,e_1(\tilde G_i(\beta))
\end{equation}
to leading order in $\ell$, which recovers Eq. \eqref{eq:ld-1} when $\tilde g(\vec \gamma) = \frac12( \delta^d( \vec \gamma-\vec \beta ) + \delta^d( \vec \gamma+\vec \beta ))$.

% \begin{equation}
    % S_{QG} = \left(\frac{k_F}{\pi}\right)^d a \ell^{d-1} \sum_{i=1}^d \sum_{j=1}^\infty e_1\left( \sum_{\beta_i \geq \frac{a}{2}j} \frac{\tilde g(\vec \beta)}{2}\right).
% \end{equation} 
 
\section{1D: Su-Schrieffer-Heeger model}

\label{sec:ssh}
\begin{figure*}
  \includegraphics[width=\textwidth]{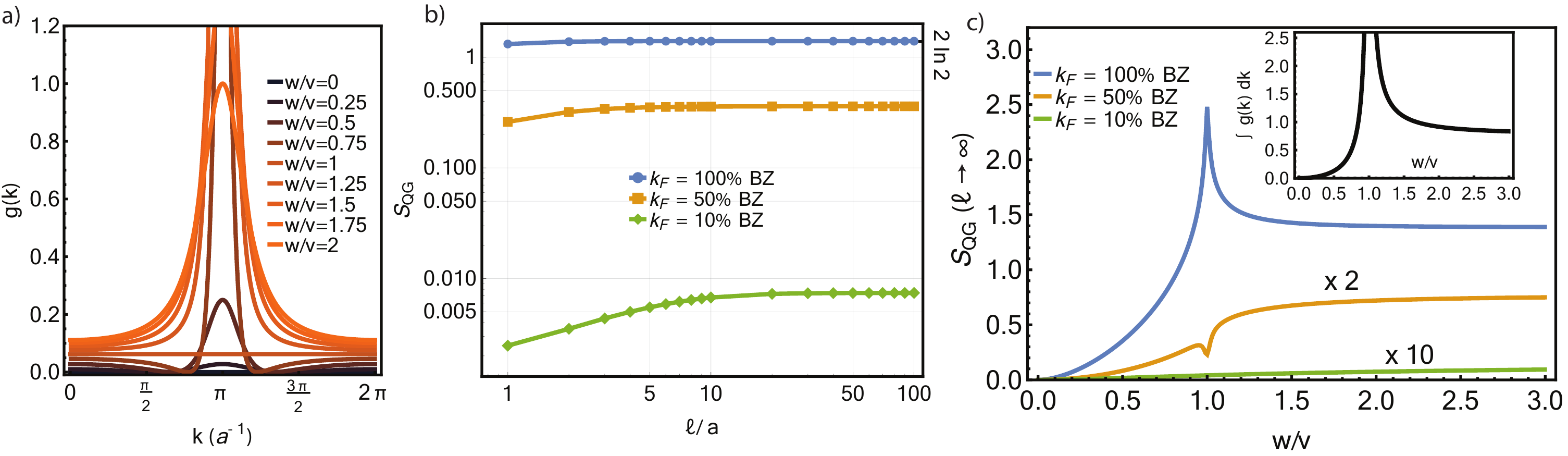}
  \caption{\textit{SSH model. } (a) Quantum metric as a function of $w/v$. The gap closes at $w=v, k= \pi/a$ and $w>v$ is topological. (b) Quantum geometric entanglement entropy $S_{QG}$ when the Fermi sea is $[0,k_F]$, for various $k_F$. It saturates to an area law at large $\ell$. (c) $S_{QG}$ at large $\ell$ versus $w/v$. Inset: integral of quantum metric over BZ versus $w/v$. System size = 600, $v=1$.}
  \label{fig:SSH}
\end{figure*}

In this section, we study the role of quantum geometry on entanglement in the Su-Schrieffer-Heeger (SSH) model \cite{Su1979Jun} at partial fillings. The SSH model describes electrons hopping on a chain with staggered hopping amplitudes without interactions. Denoting the sublattices $A$ and $B$, the Hamiltonian is
\begin{equation}
    H = \sum_j v c_{Aj}^\dagger c_{Bj} + w c_{Bj}^\dagger c_{A,j+1} + h.c.
\end{equation}
which can be written in a momentum basis as $H = \sum_k  c_{ k}^\dagger  H_k  c_{ k}$ with
\begin{equation}
    H_k =\vec d_k \cdot \vec \sigma,\quad \vec d_k = \left(v+w\cos k, w\sin k,0\right).
\end{equation}

The two bands have energies $\pm d_k$. The winding number of $ \vec d_k / d_k$ around the origin separates two phases. For $v<w$ the winding number is 0 and electrons are in a topologically trivial phase. For $v>w$, the winding number is 1 and electrons are in a topological phase, indicated by edge modes on a finite chain and a nonzero Zak phase.
\par 
Being one dimensional, the SSH chain has no Berry curvature and a single-component quantum metric. We recall that the quantum metric can be obtained from 
\begin{equation}
    1- |\left\langle u_\k|u_{\k+d\k}\right\rangle|^2 = g_{\mu\nu}(\k) dk_{\mu}dk_{\nu} + O(dk^4). 
\end{equation}
The metric in the SSH model for both bands is
\begin{equation}
    g(k) = \frac{(\vec d_k \times \vec d_k')^2}{4 d_k^4}.
\end{equation}
We plot it in Fig. \ref{fig:SSH}a. The Bloch overlaps are
\begin{equation}
    \left\langle u_k|u_q\right\rangle = \frac12 \left( 1+ \frac{(d^1_k -i d^2_k)(d^1_q +i d^2_q)}{ d_k d_q}\right). 
\end{equation}

%We would like to isolate the entanglement entropy coming purely from quantum geometry, which would not be present for free fermions with trivial Bloch overlaps (i.e. single-band fermions or fermions with no internal degrees of freedom). We do this by subtracting from $S^{(1)}$ the entropy obtained by setting $\left\langle u_k|u_q\right\rangle=1$, and we refer to this as the quantum geometric entanglement entropy $S_{QG}$. While $S^{(1)} \sim \ell^{d-1} \ln \ell$ asymptotically \cite{SwingleJul2010}, this leading term depends only on the entangling region and Fermi surface, not on the quantum geometry. Therefore $S_{QG}$ has a different asymptotic dependence on $\ell$, and our main claim is that $S_{QG} \sim \ell^{d-1}$. \par 

Since $S_{QG}$ depends on the structure of wavefunction overlaps and not directly on the band energies, we do not restrict our choice of Fermi seas to only those allowed by energetics. Rather, we simply use $\Gamma = [0,k_F]$ in this case. Alternatively, we could imagine deforming the energy bands such that $\Gamma$ is the true Fermi sea, while keeping the wavefunctions the same.
\par

We compute $S_{QG}$ by explicitly diagonalizing the overlap matrices. In Fig. \ref{fig:SSH}b, we see that $S_{QG}$ saturates to a constant for a variety of Fermi sea sizes. For a filled lower band, $S_{QG}$ approaches $\ln 2$ per boundary \cite{Ryu2006Jun}. This is a manifestation of bulk-edge correspondence \cite{Swingle2012Jul}, which relates the entanglement cut to a physical cut; the edge modes of a physical cut correspond to modes in the entanglement spectrum which contribute $\ln 2$ to the entropy.

\par We plot the dependence of $S_{QG}$ on $w/v$ in Fig. \ref{fig:SSH}c. When the first band is filled, $S_{QG}$ and $\int g(k)$ both increase monotonically up to their maximum at $w/v=1$ and then decrease monotonically, as shown in Fig. \ref{fig:SSH}c (inset). A partial understanding of this can come from the fact that the localization length of the maximally localized Wannier orbital in 1 dimension is \cite{Marzari1997Nov}
\begin{equation}
    \Omega = \int_{\text{BZ}} g(k) \,dk.
\end{equation}
Since entanglement entropy is invariant under the unitary transformation from momentum eigenstates to Wannier orbitals, one is free to choose the basis of maximally localized Wannier orbitals in computing entanglement entropy across $A$. This choice suggests that $S_{QG}$ is roughly proportional to the number of Wannier orbitals straddling the boundary of $A$, implying area-law growth proportional to $\Omega$. We discuss this point of view further in Sec. \ref{sec:wannier}.

\section{2D: Dirac cone and Chern bands}

\label{sec:2d}

\begin{figure*}
  \includegraphics[width=\textwidth]{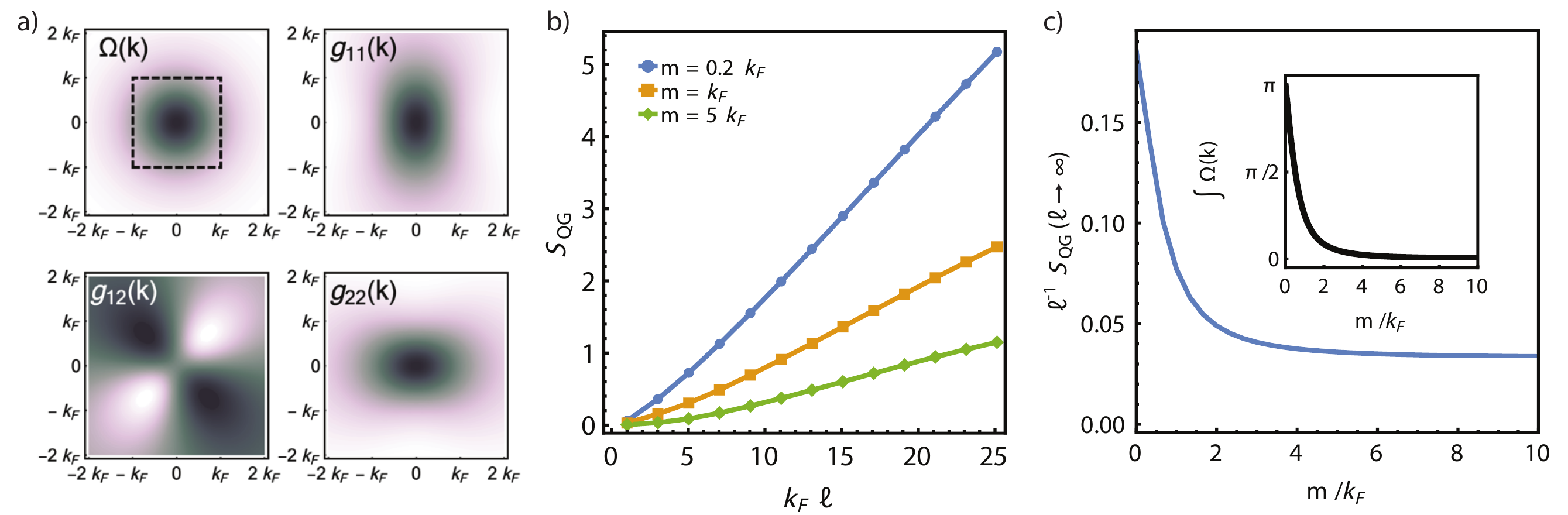}
  \caption{\textit{2D massive Dirac fermion. } (a) Density plots in $(k_x,k_y)$ of Berry curvature and quantum metric for $m=k_F$. States lying in the dashed box are filled. (b) The resulting $S_{QG}$ is linear in $\ell$ for large $\ell$ and various $m$. (c) $m$ dependence of the asymptotic $S_{QG} / \ell$. Inset: $m$ dependence of integral of Berry curvature over the Fermi sea. We note that $\Omega(\k)/2 = \sqrt{\det g(\k)}$ for the Dirac fermion model. We sampled $16^2$ $k$-points.}
  \label{fig:dirac}
\end{figure*}

\begin{figure}
\includegraphics[width=0.9\linewidth]{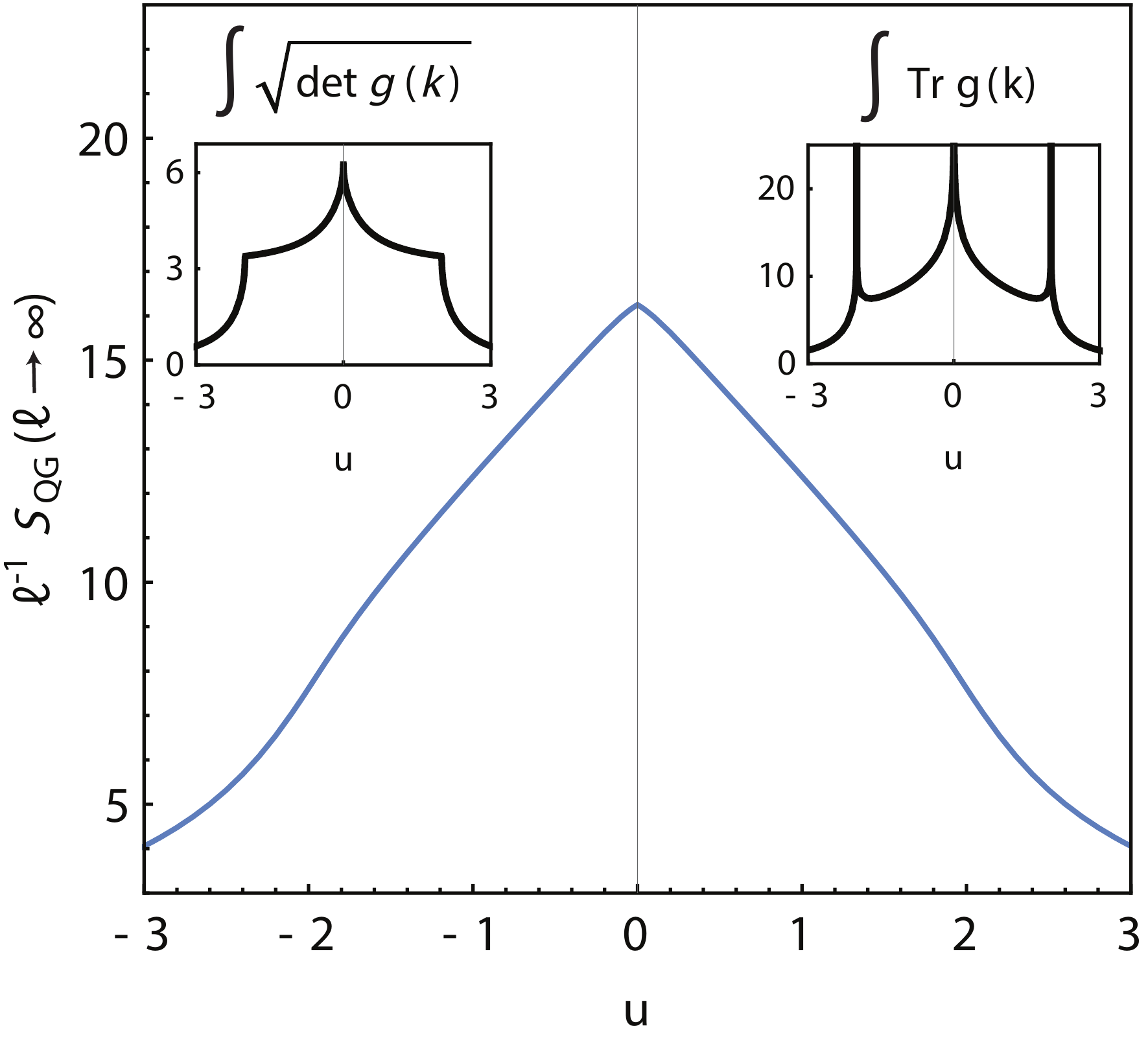}
  \caption{\textit{QWZ model.} Asymptotic $S_{QG}/ \ell$ as a function of $u$ for a filled lower band. There are topological transitions at $u=0,\pm 2$. Left inset: integral of $\surd\det g(k)$ over lower band. Right inset: integral of $\Tr g(k)$, showing logarithmic divergence at $u=0, \pm 1$, consistent with Ref. \cite{Thonhauser2006Dec}. We sampled $16^2$ $k$-points. }
  \label{fig:QWZ}
\end{figure}

Next, we study the quantum geometric entanglement entropy in two dimensional models. The simplest starting point is the massive Dirac fermion, 
\begin{equation}\label{eq:dksigma}
    H = \vec d_{\vec k} \cdot \vec \sigma, \quad \vec d_{\vec k} = (k_x, k_y, m).
\end{equation}
The Berry curvature of the lower band is
\begin{equation}
    \Omega(\k) = \frac{m}{2(k^2+m^2)^{3/2}},
\end{equation}
and the quantum metric satisfies $\Omega(\k)/2 =  \sqrt{\det g(\k)}$. We plot $\Omega$ and $g_{\mu\nu}$ in Fig. \ref{fig:dirac}a for a helpful visual. We take $\Gamma = [-k_F,k_F]^2$ to be the Fermi sea (again, we do not restrict the Fermi seas to only those allowed by energetics). Integrating $\Omega(\k)$ over the lower band gives a Berry flux of $\pi$, which is half that of a $C=1$ Chern band. We will control the quantum geometry with the dimensionless parameter $m/k_F$, since decreasing this concentrates the quantum geometry near the origin. \par 

Fig. \ref{fig:dirac}b illustrates the linear in $\ell$ behavior of $S_{QG}$ for various masses, as expected for the area law in two dimensions. In Fig. \ref{fig:dirac}c we show that the asymptotic entropy decreases monotonically with $m/k_F$. Alternatively, entropy increases monotonically with the Berry flux (or quantum volume, $\int \sqrt{\det g}$) enclosed by the Fermi surface, as shown by the inset. As before with the SSH model, we speculate that $S_{QG}$ is one of several probes quantifying ``how much quantum geometry" is enclosed by the Fermi surface. \par

The massive Dirac model describes the vicinity of the band inversion point in a topological transition. It is a continuum model, so it will be instructive to also consider a lattice model with Chern bands to illustrate the behavior of $S_{QG}$ across a topological transition. For this we use the Qi-Wu-Zhang (QWZ) model \cite{Qi2006Aug}, which takes the form of Eq. \eqref{eq:dksigma} with
\begin{equation}
 \vec d_\k = (\sin k_x,\sin k_y,u+ \cos k_x + \cos k_y).
\end{equation}
The Chern number $C$ of the lower band is
\begin{equation}
    C = \begin{cases} 0 & |u| >2\\
    -1 &-2 < u < 0\\
    1 & 0 < u < 2
    \end{cases}.
\end{equation}

At the transition at $u=-2$ ($u=2$), a Dirac point occurs at $\k = 0$ ($\k = (\pi,\pi)$). At the transition at $u=0$, a pair of Dirac points occur at $\k = (0,\pi)$ and $(\pi,0)$. In order to avoid effects due to specific Dirac points, we choose the Fermi sea to be the entire lower band. As expected for a 2D model, $S_{QG}\sim \ell $ for large $\ell$. In Fig. \ref{fig:QWZ} we plot the dependence of $\lim_{\ell\to \infty} S_{QG}/\ell$ on $u$. In contrast with the $m=0$ behavior of $S_{QG}$ for the massive Dirac fermion, the entropy for the QWZ model does not show sharp features across the band-touching points. This is also in contrast with the integrals of $\sqrt{\det g(k)}$ and $\Tr g(k)$ over the lower band (insets); the former shows cusp-like transitions while the latter diverges at $u=\pm 2,0$. $S_{QG}$ is not directly proportional to either of these quantities, suggesting again that $S_{QG}$ is a quantum geometric probe which goes beyond leading order in $dk$.

\section{Experimental measures}
\label{sec:exp}

Although entanglement entropy has proven quite useful conceptually in condensed matter physics, it is difficult to measure directly. This is in part because it is nonlinear in $\rho$ and highly nonlocal. In the context of optical lattices, Refs. \cite{Pichler2013Jun} and \cite{Daley2012Jul} proposed schemes for measuring the entanglement entropy using quantum interference of two copies of the many-body state. Using this method, Ref. \cite{Islam2015Dec} measured the 2nd R\'enyi entropy in the ground state of the four-site Bose-Hubbard model in both the Mott insulating and superfluid phases. In general, one can interfere $n$ copies of the state to measure an $n$th-order polynomial in $\rho$ \cite{Ekert2002May}. \par

Such a protocol is generally impractical for solid state devices. Along different lines, Ref. \cite{Klich2009Mar} suggested particle number fluctuations as an effective measure of many-body entanglement. This is based on the fact that fermion entanglement entropy has an exact expansion in terms of even cumulants $V^{(2j)}_A$ of the particle number distribution which rapidly converges at large $N$ \cite{Calabrese2012Apr}. In particular, 

\begin{equation}\label{eq:Vseries}
    S = \frac{\pi^2}{3} V^{(2)} + \frac{\pi^4}{45} V^{(4)}+ \frac{2\pi^6}{945} V^{(6)} + \cdots
\end{equation}
where $V^{(2)} = \Tr \mathbb{A}(1-\mathbb{A})$ and similar expressions exist for the other cumulants. Therefore the fluctuations of current flowing through a quantum point contact serve as a good probe of the many-body entanglement, and thus of quantum geometry. \par 

Let us work out the dependence of the leading term on quantum geometry. We write
\begin{equation}\label{eq:Akk'}
    \mathbb{A}_{\k \k'} = I_{\k-\k'} \left\langle u_\k | u_{\k'}\right\rangle
\end{equation}
where $I_{\k-\k'} = L^{-d} \int_A d\x \,\, e^{- i (\k-\k') \cdot \x}$ . As before, we use a cubic region $A$ with volume $\ell^d$. The particle number variance can be written as 
\begin{align}
    V^{(2)} &= N (\ell/L)^d - \sum_{\k\k'} |A_{\k\k'}|^2\nonumber \\
    &= N(\ell/L)^d - 2^{-d}\sum_{\p} |I_{\p}|^2 \sum_\k|\left\langle u_\k|u_{\k+\p}\right\rangle|^2
\end{align}
letting $\k' = \k + \p$. When $\left\langle u_\k | u_{\k'}\right\rangle = 1$, an explicit calculation using the Widom conjecture gives leading $\ell^{d-1} \ln \ell$ behavior \cite{Calabrese2012Apr}. As with entropy, we define the \textit{quantum geometric cumulant} $V_{QG}^{(m)}$ by subtracting the result when $\left\langle u_\k | u_{\k'}\right\rangle = 1$.

\par 

$|I_\p|^2$ is strongly peaked near $\p= 0$, so this motivates an expansion
\begin{equation}
    1-|\left\langle u_\k|u_{\k+\p}\right\rangle|^2  = \sum_{j=1}^\infty \Omega_{\mu_1,\ldots, \mu_{2j}}^{(2j)}(\k)\,  p_{\mu_1} \cdots p_{\mu_{2j}}
\end{equation}
in even powers of momentum, where $\Omega^{(2j)}_{\mu_1,\ldots,\mu_{2j}}$ is a symmetric rank-$2j$ tensor. For instance, $\Omega^{(2)}_{\mu_1\mu_2} = g_{\mu_1\mu_2}(\k)$, the quantum metric. Let us study the contribution at order $2j$:
\begin{equation}
    \delta_{2j} V_{QG}^{(2)} = 2^{-d}\sum_\p |I_\p|^2 p_{\mu_1}\cdots p_{\mu_{2j}} \sum_\k \Omega^{(2j)}_{\mu_1\ldots\mu_{2j}}(\k).
\end{equation}
As before, let's assume a Fermi sea $\Gamma = [-k_F, k_F]^d$; the main results are unaffected by this choice. The sum over $\p$ vanishes if an odd power of $p_i$ occurs. Moreover, the contribution which is leading in $\ell$ comes from $\mu_1 = \cdots = \mu_{2j}$. Letting 
\begin{equation}
    \Omega^{(2j)}(\k) \equiv \sum_{\mu} \Omega_{\mu \cdots \mu}^{(2j)}(\k)
\end{equation}
and converting the sums to integrals, we may write
\begin{align}
    &\delta_{2j} V_{QG}^{(2)} \nonumber\\
    &=\frac{L^{2d}}{2^dd} \int_{2\Gamma} \frac{d^d\p}{(2\pi)^d}  \, |I_\p|^2 p^{2j} \int_\Gamma \frac{d^d \k}{(2\pi)^d}  \Omega^{(2j)}(\k)+ O(\ell^{d-2})\nonumber\\
     &= \frac{2^{2j-d}}{(2j-1)\pi}k_F^{2j-1} \ell^{d-1} \int_\Gamma \frac{d^d \k}{(2\pi)^d} \Omega^{(2j)}(\k) + O(\ell^{d-2}).
\end{align}
Altogether, 

\begin{equation}
    V_{QG}^{(2)} = \bar \Omega \ell^{d-1} + O(\ell^{d-2})
\end{equation}
where 
\begin{equation}
    \bar \Omega = \sum_{j=1}^{\infty} \frac{2^{2j-d}k_F^{2j-1}}{(2j-1)\pi} \int_\Gamma \frac{d^d\k}{(2\pi)^d} \Omega^{(2j)}(\k).
\end{equation}
Like the quantum geometric entanglement entropy, $V_{QG}^{(2)}$ follows an area law. Moreover, if $\Gamma$ is small compared to the Brillouin zone, the quantum metric term dominates. In other words

\begin{equation}\label{eq:probe}
    V_{QG}^{(2)} \approx 2^{2-d} (k_F/\pi) \ell^{d-1} \int_\Gamma \frac{d^d\k}{(2\pi)^d} \Tr g(\k)
\end{equation}
for small $k_F$ and large $\ell$. This allows for a direct, practical probe of the quantum metric: by slightly doping a band and measuring the particle number variance at large $\ell$, one directly measures $\Tr g(\k)$ near the band bottom.\par 

To conclude, particle number fluctuations give an experimental measure of the entanglement entropy via Eq. \eqref{eq:Vseries}, which converges rapidly at large $N$. In an experiment, the leading $\ell^{d-1} \ln \ell$ behavior must be subtracted to extract the quantum geometric contribution. We confirmed that the quantum geometric contribution to particle number variance follows an area law (like $S_{QG}$) and increases with quantum geometry, and we arrived at a simple probe of the quantum metric near the band bottom, Eq. \eqref{eq:probe}.

\section{Relation to Wannier spread}
\label{sec:wannier}

A partial understanding of the relation between quantum geometric entanglement entropy and quantum geometric quantities such as $\int \Tr g(\k)$ can come from considering maximally localized Wannier orbitals. The Wannier spread (taken to mean the real-space variance of a Wannier orbital) for a single isolated non-topological band can be written as  
\begin{equation}
    \Omega = \Omega_I + \tilde \Omega
\end{equation}
where we've adopted the notation of Ref. \cite{Marzari1997Nov}. Here
\begin{equation}
    \Omega_I = V\int_{\text{BZ}} \frac{d^d\k}{(2\pi)^d} \Tr g(\k)
\end{equation}
is a gauge-invariant piece depending on the quantum metric and $\tilde \Omega$ depends on the Berry connection: 
\begin{equation}\label{eq:tildeomega}
    \tilde \Omega = V \int_{\text{BZ}} \frac{d^d\k}{(2\pi)^d} |\vec A_\k - \r|^2
\end{equation}
where $\r = V \int_{BZ} \frac{d^d\k}{(2\pi)^d} \vec A_\k$. The gauge choice which minimizes $\tilde \Omega$ (and thus $\Omega$) is the transverse gauge $\vec \nabla_\k \cdot \vec A_\k = 0$ \cite{Marzari1997Nov}. For $d=2$ or $3$, this is equivalent to choosing the Berry connection by solving the Laplace equation
\begin{equation}\label{eq:laplacian}
    \nabla^2_\k \vec A_\k = - \vec \nabla_\k \times \vec \Omega_\k
\end{equation}
where $\vec \Omega_\k = \vec \nabla_\k \times \vec A_\k$ is the Berry curvature, with suitable generalizations for other $d$. We emphasize that while Eq. \eqref{eq:tildeomega} is gauge-dependent, its \textit{minimum} is gauge-invariant and depends directly on the Berry curvature. \par

Entanglement entropy across a region $A$ can be thought of as a measure of the number of modes straddling the boundary of $A$. It is gauge-invariant and invariant under unitary rotations in the Hilbert subspace of occupied states. For a filled non-topological band, one is free to consider the basis of maximally localized Wannier orbitals, in which case one may speculate that $S_{QG}$ should be area law and roughly proportional to $\Omega$ for the maximally localized Wannier orbitals, which is defined entirely in terms of gauge-invariant quantum geometric data. \par 
We remark, however, that in the previous sections we have provided several counterexamples to an exact $S_{QG} \sim \Omega \ell^{d-1}$ scaling relation. The exact expression of $S_{QG}$ in terms of quantum geometric data is an interesting problem for future research. \par 

The picture discussed above breaks down for topological bands and fractional band filling. For topological bands, there is no basis of exponentially localized Wannier orbitals, and such a basis generally mixes states of all momenta. However, the previous sections have shown that an area law scaling of $S_{QG}$ holds in these cases as well.

%Finally, we remark that Ref. \cite{Brouder2007Jan} noted that the more regular the Bloch states $u_\k$ are, the more localized the Wannier functions. This is consistent with our observations of the role of quantum geometry (which increases as the $u_\k$ become less regular) in entanglement entropy, and making this precise would be an interesting future direction. 

\section{Discussion}
\label{sec:discussion}

To summarize, we have introduced the \textit{quantum geometric entanglement entropy} $S_{QG}$ which captures the contribution of quantum geometry to real-space von Neumann entanglement entropy for noninteracting fermion systems. For the case of uniform quantum geometry an explicit form was given, and shown to be area-law, asymptotically in the linear size of the entangling region. We also demonstrated the area-law behavior for the SSH model, the 2D massive Dirac cone, and the QWZ model. $S_{QG}$ was shown to resemble, but not precisely match, various quantum geometric quantities such as $\int \Tr g(k)$, which in turn is related to the minimal Wannier orbital spread for a non-topological band. Evidently $S_{QG}$ is a refined measure of quantum geometry which is sensitive to all orders in momentum separation, unlike the quantum metric and Berry curvature, which are second-order in derivatives. Finally, we related $S_{QG}$ to particle number fluctuations, which are measurable using a quantum point contact. A corollary was a direct way to measure the quantum metric at the bottom of a band using the particle number variance. \par 

One interesting problem for future work is the explicit formula for the leading behavior of $S_{QG}$ as a functional of $g(\k,\q) \equiv \left\langle u_\k|u_\q\right\rangle$ (we have completed the case $g(\k,\q) = g(\k-\q)$). The general case would amount to an analytical calculation of $S_{QG}$ for all models where the Bloch states are known.\par

The interplay of weak interactions and quantum geometry from an entanglement viewpoint is also an interesting topic for future study. For example, what are the effects of underlying quantum geometry on the entanglement properties of a Fermi liquid? In this case, both real or momentum-space entanglement cuts could give nontrivial information \cite{Flynn2022Mar}. \par

\textit{Acknowledgments--} I'd like to thank Liang Fu for his support and encouragement during this work. I'd also like to thank Phil Crowley, Trithep Devakul, Aidan Reddy, and Senthil Todadri for helpful conversations. 
%I would like to thank Phil Crowley, Trithep Devakul, and Senthil Todadri for helpful comments. I would like to thank Liang Fu for support and encouragement during this work.

\par

\bibliographystyle{apsrev4-1}
\bibliography{bib}

\appendix
\clearpage

\end{document}